Systematic Comparison of Eight Substrates in Growth of FeSe$_{0.5}$Te$_{0.5}$ Superconducting Thin Films

Yoshinori Imai[1, 4, *], Takanori Akiike[1, 4], Masafumi Hanawa[2, 4], Ichiro Tsukada[2, 4], Ataru Ichinose[2, 4], Atsutaka Maeda[1, 4], Tatsuo Hikage[3], Takahiko Kawaguchi[3, 4], and Hiroshi Ikuta[3, 4]

[1] Department of Basic Science, the University of Tokyo, 3-8-1 Komaba, Meguro, Tokyo 153-8902, Japan

[2] Central Research Institute of Electric Power Industry, 2-6-1 Nagasaka, Yokosuka, Kanagawa 240-0196, Japan

[3] Department of Crystalline Materials Science, Nagoya University, Chikusa, Nagoya 464-8603, Japan

[4] Transformative Research-Project on Iron Pnictides (TRIP), Japan Science and Technology Agency, 5 Sanbancho, Chiyoda, Tokyo 102-0075, Japan

Abstract

We have investigated the crystal structure and superconducting properties of thin films of FeSe$_{0.5}$Te$_{0.5}$ grown on eight different substrates. Superconductivity is not correlated with the lattice mismatch, but rather it is correlated with the degree of in-plane orientation and with the lattice parameter ratio *c*/*a*. The best superconducting properties are observed in films on MgO and LaAlO$_3$ ($T_c^{\text{zero}}$ of 9.5 K). TEM observation shows that the presence or absence of the amorphous-like layer at the substrate surface plays a key role in determining the structural and superconducting properties of the grown films.

* E-mail address: imai@maeda1.c.u-tokyo.ac.jp

After the discovery of superconductivity in F-doped LaFeAsO,[1] numerous studies on iron-based superconductors have been carried out. One common iron-based superconductor is FeSe with a superconducting transition temperature $T_c$ of 8 K,[2] and the partial substitution of Te for Se raises $T_c$ to a maximum of 14 K.[3] This material has the tetragonal PbO-type structure, which is the simplest structure of all the iron-based superconductors. Thus, FeSe and related materials are considered the most suitable systems to investigate how superconductivity correlates to the crystal structures. Many studies on the film growth of $FeSe_{1-x}Te_x$ have already been reported.[4–11] However, the question of what substrates are suitable for the growth of thin $FeSe_{1-x}Te_x$ films remains controversial. For example, Kumary et al.[6] reported that the $T_c$ value of the film on SrTiO$_3$ (STO) was higher than on LaAlO$_3$ (LAO). In contrast, Han et al.[7] reported an opposite result; the $T_c$ value on LAO was higher than that on STO. In addition, Bellingeri et al.[8] reported that there was no obvious difference in the $T_c$ values of films grown on STO and LAO when their thickness exceeded 200 nm. Recently, we have succeeded in growing thin superconducting $FeSe_{0.5}Te_{0.5}$ films on MgO and LaSrAlO$_4$.[10] A strong correlation has been found between the $c$-axis length and $T_c$ depending on the substrate materials, but it cannot be explained by a difference in the lattice mismatch between films and substrates. In this article, we present a systematic comparison of the crystal structures of thin films of FeSe$_{0.5}$Te$_{0.5}$ grown on eight different substrates, and discuss the influence of the substrate on the superconductivity of films.

Thin films were grown under vacuum ( ~ 10$^{-5}$ Torr) by pulsed laser deposition (PLD) using a KrF laser (repetition rate: 10 Hz).[10] FeSe$_{0.5}$Te$_{0.5}$ polycrystalline pellets were used as the targets. We used eight kinds of single-crystal substrate: MgO ($a$ = 4.212 Å), LaAlO$_3$ (LAO, $a$ = 3.790 Å), SrTiO$_3$ (STO, $a$ = 3.905 Å), LaSrAlO$_4$ (LSAO, $a$ = 3.754 Å), LaSrGaO$_4$ (LSGO, $a$ = 3.844 Å), yttria stabilized zirconia Y:ZrO$_2$ (YSZ, $a/\sqrt{2}$ = 3.624 Å), (LaAl)$_{0.7}$(SrAl$_{0.5}$Ta$_{0.5}$)$_{0.3}$O$_3$ (LSAT, $a/2$ = 3.869 Å), and Al$_2$O$_3$ ($a$ = 4.758 Å). We use the (001) plane of LSAO and LSGO, the (0001) plane of hexagonal Al$_2$O$_3$, and the (100) plane of other crystals. To isolate the effect of the substrate materials on superconductivity of films, it is of crucial importance to keep the growth conditions during deposition identical. The substrate temperature and the film thickness were fixed at 573 K and at approximately 50 nm, respectively. The crystal structure and the orientation of the films were characterized by a θ-2θ and a 4-circle X-ray diffraction (XRD) using Cu Kα radiation at room temperature. We also performed a transmission electron microscopy (TEM) observation. The electrical resistivity (ρ) was measured by a four-terminal method from 2 to 300 K.

Figure 1 shows the XRD patterns of the eight films. Except for a few unidentified peaks, only the 00$l$ reflections of a tetragonal PbO-type structure are observed, which shows that the out-of-plane alignment is excellent. It should be noted that the *c*-axis orientation is observed even in the film prepared on the (0001) plane in hexagonal $Al_2O_3$, as shown in Fig. 1 (h). This indicates that the $FeSe_{1-x}Te_x$ films intrinsically favor two-dimensional growth. The temperature dependence of ρ is summarized in Fig. 2. As can be easily seen, the eight films exhibit a variety of ρ(T) behavior. Except for the film on $Al_2O_3$, the ρ value of these films at $T = 300$ K is 0.35-0.70 mΩcm, which is as small as that of the single-crystal $FeSe_{0.5}Te_{0.5}$.[12] The films on MgO, LAO, and STO show zero resistivity at 9.2 K, 9.4 K, and 6.6 K, respectively. The $T_c$ values of the films on MgO and LAO are as high as those of polycrystalline targets, while the $T_c$ value of the film on STO is slightly lower than those of the other two films. The films on LSAO and LSGO, in contrast, exhibit only the onset of superconducting transition at about 4 K, but do not show zero resistivity down to 2 K. These two films exhibit the crossover from semiconducting to metallic behavior before the superconducting transition with decreasing temperature. In the case of films on the other three substrates, no metallic behavior is observed in the temperature dependence of ρ.

To identify the reason for the difference in the superconducting properties, we investigate the in-plane orientation. Figures 3 (a)-(e) show the φ scans of 101 peak from the five films. Figures 3 (a)-(c) clearly show that the films on MgO, LAO, and STO have a four-fold in-plane orientation. The full width at half maximum (FWHM) is Δφ ~ 2.34°, 0.65°, and 1.89° for the films on MgO, LAO, and STO, respectively. In contrast, in the film on LSAO (and also LSGO), the intensity is much smaller than in the former three films, and the peak is broad, as shown in Fig. 3 (d), where FWHM is Δφ ~ 7.50°. No clear evidence of the in-plane alignment is observed in the other three films, as shown in Fig. 3 (e).

To clarify the substrate-specific difference, the $a$- and $c$-axis lengths of grown films, $T_c^{onset}$ (defined as the temperature where the electrical resistivity deviates from the normal state behavior), and $T_c^{zero}$ are tentatively plotted as functions of the ratio of $c$ divided by $a$, $c/a$, in Fig. 4. The values in brackets are the calculated lattice misfit, $M$, which is defined as $M = (a_{sub}-a_0)/a_{sub} \times 100$, where $a_0 = 3.7947$ Å[3] and $a_{sub}$ is the above-mentioned $a$-axis length of the substrate. First, there is no simple relation between the $a$-axis lengths of the films and those of the substrates. For example, the $a$-axis length of the film on LAO, which has a vanishingly small lattice misfit (−0.12%), is quite similar to that of the film on MgO, which has the largest misfit (9.9%) of all substrates investigated. It is also interesting to see that the film on YSZ, which has the smallest $a$-axis length among all substrates, has the largest $a$-axis length of all the films. Second, there is no correlation between superconducting properties and the lattice misfit. This is clear because the $T_c$ values of the films on MgO and LAO are almost the same, in spite of the big difference in the lattice misfit between these two substrates. We find that $T_c$ can be scaled by the ratio $c/a$, as is clearly seen in Figs. 4(c) and (d). The films with comparably high values of $T_c$, that is, the films on MgO, LAO, and STO, have large values of $c/a$. Recall that these three films generally have a high in-plane quality, as was already shown in Figs. 3 (a)-(c). Therefore, these results strongly suggest that superconducting properties of $FeSe_{0.5}Te_{0.5}$ films are affected by the ratio $c/a$ and by the degree of the in-plane orientation. Margadonna et al.[13] observed that $T_c$ increases with the ratio $(a + b)/2c$ at high pressure in polycrystalline FeSe. Their results are in contrast to our results in $FeSe_{0.5}Te_{0.5}$ films. Understanding the origin of this difference will be the subject of a future study.

Finally, to understand the reasons why the $a$-axis lengths of grown films and those of the substrates are not correlated and why the degree of in-plane orientation depends strongly on the substrate materials, we show the high-magnification TEM images of films on LAO and YSZ in Figs. 3 (f)-(i). In the case of LAO (Figs. 3 (f) and (g)), the interface between the film and the substrate is smooth, and the film grows just on the substrate.  In contrast, an amorphous layer is likely to exist between the grown film and the substrate in the film on YSZ, which is indicated by the up arrows in Figs. 3(h) and (i).  Thus, the $a$-axis length and the orientation of the substrate material are not reflected in the grown film because of the presence of the amorphous-like layers.  It should be noted that such amorphous-like layers commonly exist in the films with poor or no superconductivity.  Therefore, we conclude that the presence of the amorphous-like layer plays a key role in the structural and superconducting properties of the grown films.  We believe that this is the reason why the lattice mismatch does not play an important role in the growth of $FeSe_{0.5}Te_{0.5}$ films.

In conclusion, we have successfully grown the $FeSe_{1-x}Te_x$ films on eight different substrates by the PLD method.  The superconductivity did not correlate with the lattice mismatch, but rather correlated with the degree of in-plane orientation and with the lattice parameter ratio $c/a$.  The best superconducting properties are observed in films on MgO and $LaAlO_3$ ($T_c^{zero}$ of 9.5 K), among the substrate materials investigated in this study.  TEM observation shows that the presence or absence of an amorphous-like layer at the substrate surface plays a key role in determining the structural and superconducting properties of the grown films.  Thus, the epitaxial growth of thin films of FeSe$_{0.5}$Te$_{0.5}$ is nontrivial.

List of Figures

FIG. 1. (color online) XRD patterns of films grown on eight different substrates. The asterisks represent the peaks resulting from the substrate, and the sharps represent the unidentified peaks.

FIG. 2. (color online) Temperature dependence of the electrical resistivity for eight films.

FIG. 3. (color online) XRD in-plane patterns ($\varphi$ scans) of the 101 peak from FeSe$_{0.5}$Te$_{0.5}$ films on (a) MgO, (b) LAO, (c) STO, (d) LSAO, and (e) YSZ, and high-magnification TEM cross sectional images of FeSe$_{0.5}$Te$_{0.5}$ films on LAO [(f) Overall view (g) Enlarged view] and YSZ [(h) Overall view (i) Enlarged view].

FIG. 4. (color online) Dependence of the (a) *a*-axis and (b) *c*-axis lengths of grown films, (c) $T_c^{onset}$, and (d) $T_c^{zero}$ on the ratio of *c* divided by *a*.

Fig. 1

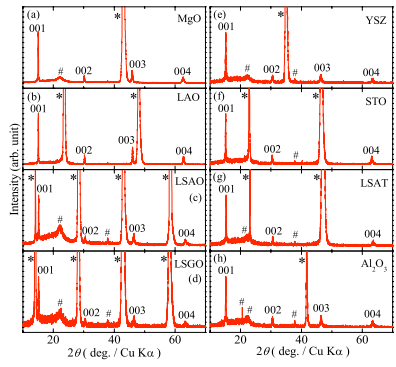

Fig. 2

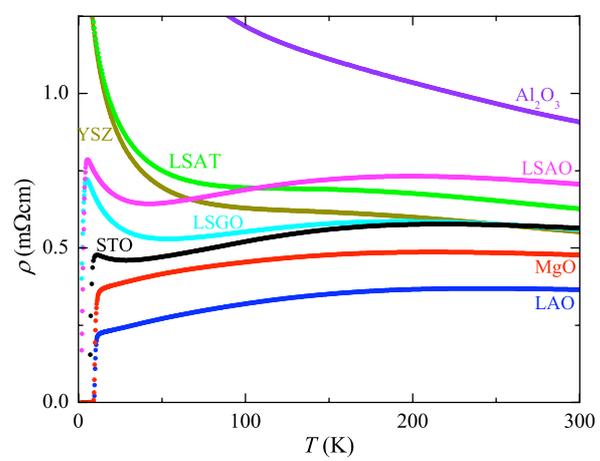

Fig. 3

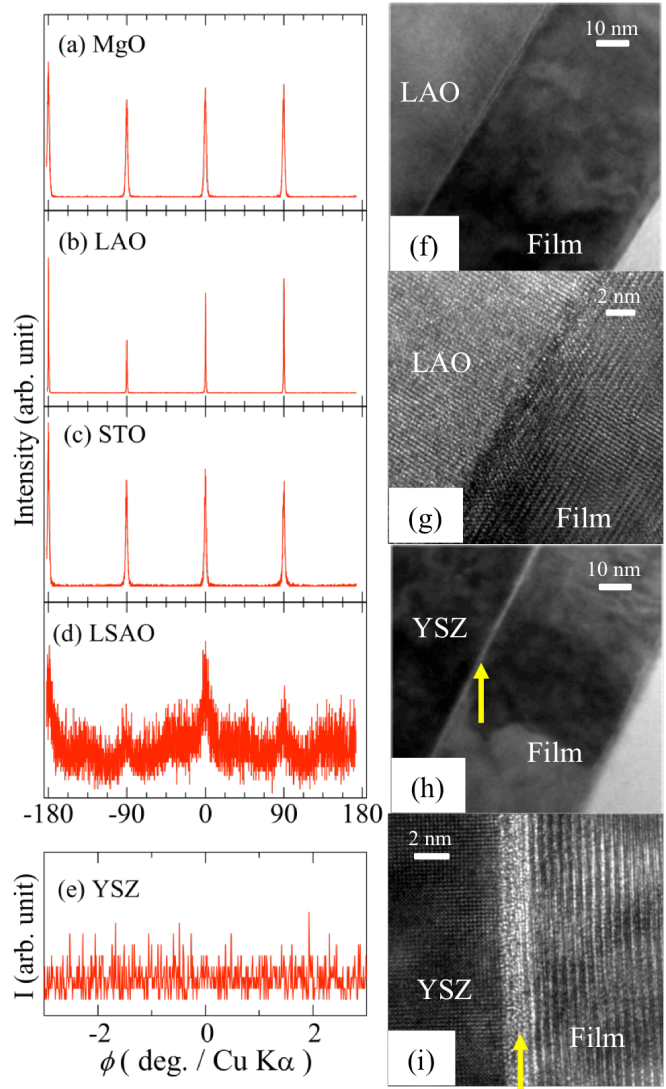

Fig. 4

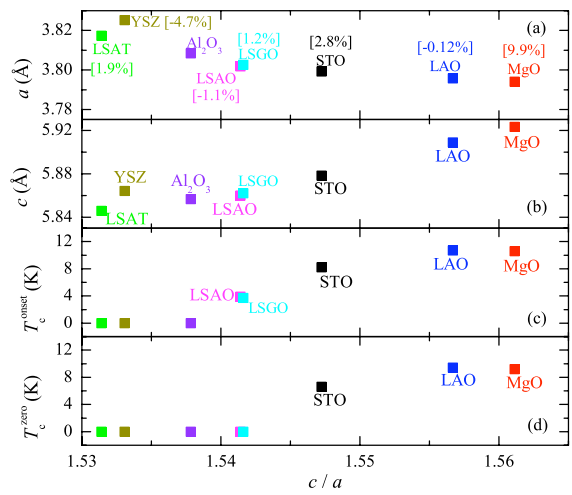